\documentclass{article}

\usepackage{PRIMEarxiv}

\usepackage[utf8]{inputenc} 
\usepackage[T1]{fontenc}    
\usepackage{hyperref}       
\usepackage{url}            
\usepackage{fancyhdr}       
\usepackage{graphicx}       
\usepackage{adjustbox}      
\usepackage{array}          
\usepackage{microtype}      

\graphicspath{{figures/}}     

\title{Ten simple rules for collaborating with wet lab researchers for computational researchers}

\author{
\textbf{Mark D. Robinson}$^{1,2,3,4}$, 
\textbf{Peiying Cai}$^{1,2,4}$,
\textbf{Martin Emons}$^{1,2,4}$,
\textbf{Reto Gerber}$^{1,2,4}$,\\
\textbf{Pierre-Luc Germain}$^{1,2,4,5}$,
\textbf{Samuel Gunz$^{1,2,4}$}, 
\textbf{Siyuan Luo$^{1,2,4}$},
\textbf{Giulia Moro$^{1,4}$},
\textbf{Emanuel Sonder$^{1,2,4,5}$},\\
\textbf{Anthony Sonrel$^{1,2,4}$},
\textbf{Jiayi Wang$^{1,2,4}$},
\textbf{David Wissel$^{1,2,4}$},
\textbf{Izaskun Mallona$^{1,2,3}$}\\ 
\\
$^1$ Department of Molecular Life Sciences, University of Zurich, Winterthurerstrasse 190, CH-8057 Zurich, Switzerland;\\ $^2$ SIB Swiss Institute of Bioinformatics;\\ $^3$ Co-correspondence \texttt{\{mark.robinson,izaskun.mallona\}@mls.uzh.ch}; $^4$ Equal contribution;\\
$^5$ D-HEST Institute for Neuroscience, ETH Zurich, Winterthurerstrasse 190, CH-8057 Zurich, Switzerland.\\ 
}

\begin{document} 
\maketitle

\vspace{-0.5cm}

\begin{abstract}
Computational biologists are frequently engaged in collaborative data analysis with wet lab researchers. These interdisciplinary projects, as necessary as they are to the scientific endeavour, can be surprisingly challenging due to cultural differences in operations and values. In these Ten Simple Rules guide we aim to help dry lab researchers identify sources of friction; and provide actionable tools to facilitate respectful, open, transparent and rewarding collaborations.
\end{abstract}

\vspace{0.5cm}


\section*{Introduction} 

Many computational biologists (or biologists doing applied computational analysis, or statisticians making inferences from biological data, or physicists modeling biological processes) are often faced with ``collaborative data analysis'' projects. Here, we specifically use the term collaborative data analysis instead of support or service to better represent our role in the scientific endeavour. The perspective represented here is that of a dry lab research group primarily engaged in assessing and developing new computational tools to process, interpret, explore and make inferences from complex molecular data. In theory, effective development of computational methods happens in harmony with the analysis of collaborators’ ``wild caught'' data and all the baggage that comes with cross-disciplinary multi-research-group (sometimes multi-silverback-group-leader) collaborations. In the ideal case, our contributions glue together robust streams of evidence for fascinating and complicated biological phenomena; in other cases, one could perceive such work as a rather unrewarding and pointless publish-or-perish endeavour.

\begin{figure}[h]
  \centering
  \fbox{\includegraphics[width=0.99\linewidth]{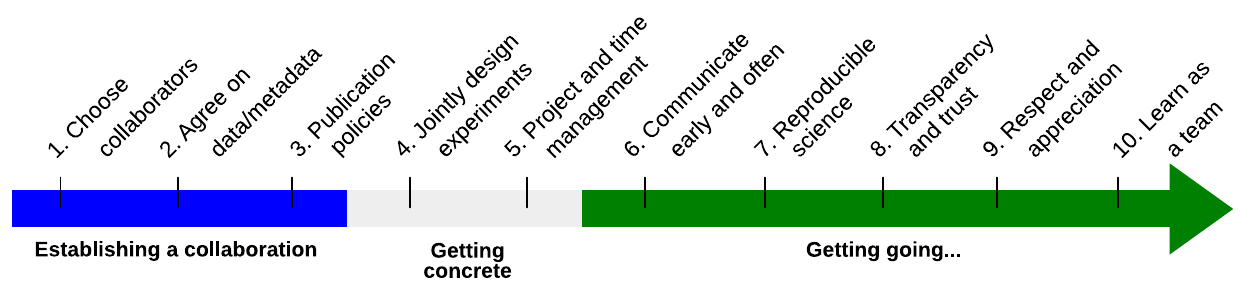}}
  \caption{Typical timeline of a collaboration between experimentalists and analysts. The different steps are coarsely grouped in three phases on the timeline, but some steps are regular processes throughout the collaboration.}
  \label{fig:figure1}
\end{figure}

Despite our primary role as methodologists and that we often spend only a minority of our time on active collaboration, this cross-culture working arrangement can sometimes be a greater source of tension than the ``blissful'' world of pure method development. This 10 simple tips guide unpacks not only the pain points, forking paths, and mundane though important logistics of everyday collaborative data analysis (Figure~\ref{fig:figure1}), but also reveals the wild and random though nonetheless inspiring adventures that a day in the life of a collaborative computational biologist brings.

\clearpage

\section*{Rule 1: Choose your collaborators wisely} 
\label{rule1_choose}

\begin{flushright}
\rightskip=1cm\textit{``It is better to be alone than in bad company''} \\
\vspace{.2em}
\rightskip=0cm -- George Washington
\end{flushright}

Our experience has been that good collaborators are those that are engaged not only in their own domain, but also have a thirst for knowledge about computational aspects. In many ways, this mirrors our interest in collaborative data analysis: we are tasked with handling data analysis aspects, but we endeavour to understand as much as possible of the biological context of the data. Less attractive collaborators are those that simply want a table of P-values or a set of figures; worst-case scenario are those that perceive that we simply press a button to do so. 

Choosing collaborators represents the first fundamental step for collaboration between wet-lab and dry-lab researchers. Without an aligned choice of collaborating labs, discrepancies in scientific interest and values are harder to navigate later on. Two questions should be considered. First, is there an adequate scientific match in the collaboration (e.g., for us, analysis of a certain type of omic data)? That is, do the groups complement each other well in terms of scientific interests and skills? Second, are (research) values between groups aligned? Meaning, do both groups have broadly similar ideas of how research is to be conducted day-to-day? For example, is excellence defined as a CNS paper, or is it robust and high quality science?

In practice, collaborations originate between researchers who know each other already (e.g., in the same institute). Thus, the scientific match is usually there or can be easily established. Ascertaining the alignment of research values is more difficult and sometimes requires months and years of working together to fully comprehend. 

In addition, choosing collaborators can also be thought of as a process that repeats itself at natural stopping points, such as the finishing of a project (stage). It can be advantageous to evaluate at regular intervals whether all parties are getting what they had hoped out of the collaboration and re-discuss the collaboration if not.

In our experience, most friction arises from having different expectations that are not clearly expressed nor negotiated. To detect and smoothen possible misalignments (for this and other rules) we suggest both teams fill an expectations form (such as Table~\ref{form}), compare them, and discuss conflicting views early on. Explicit is better than implicit. 

\section*{Rule 2: Agree on data and metadata structure} 
\label{rule2_data}

\begin{flushright}
\rightskip=1cm\textit{``With big data comes big responsibilities''} \\
\vspace{.2em}
\rightskip=0cm -- Kate Crawford
\end{flushright}

In omics data analysis, there are typically many stages to an analysis, each one having its own type of data (e.g., raw data, filtered data, normalized data, modelling, inference, etc). How and what data is shared internally between the project members during the project should be agreed upon, i.e how are data shared between collaborators, what (intermediate) data can be shared, data formats that can be easily read by all researchers. The FAIR principles \cite{wilkinson2016fair} on data sharing offer good guidelines and can be useful later when publishing the open data of the project.

Sensitive data (frequently patient-related personal data) represent a special case as they require additional protection with data storage and sharing.  A data sharing agreement should be determined between the collaborators. Analysts working on these datasets are obliged to take steps in accordance with the law and the signed agreement. Normally the agreement specifies who is granted access to the data; the purpose of the data analysis; what measures need to be taken (e.g., access, logging) to ensure the privacy and integrity of the data. 

Clear standards for metadata format and style are often lacking (even among some fields of computer science), which can often lead to misleading or erroneous data and ultimately to biased results. A good practice is to agree on a common and systematic way of sharing the metadata, which can both be easily accessible and modified by experimentalists, and easily parsed by analysts, and consistent enough to be integrated in an analysis workflow.  

Actionable items: Follow the FAIR guidelines (\url{https://www.go-fair.org/fair-principles/}); define file naming policies; be aware of sensitive data handling regulations (in Europe) \cite{shabani2019reidentifiability,party2011advice}; follow good practices for spreadsheets \cite{broman2018data}.

\section*{Rule 3: Define publication policies} 
\label{rule3_publication}

\begin{flushright}
\rightskip=1cm\textit{``The only good ideas are the ones I can take credit for''} \\
\vspace{.2em}
\rightskip=0cm -- Richard Stevens
\end{flushright}

Explicitly ask your collaborator's expectations on research dissemination, openly express yours and reach consensus. Expectations to discuss include strategies on paper publishing, conferences, and other deliverables, such as software; in particular, basic ground rules for author ordering should be broached.

For academic papers, discuss potential target and no-go journals, (posting and updating) preprints, and open access. For other deliverables, particularly software and data analysis workflows, discuss intellectual property, copyright, open software licenses, and whether these deliverables can be disseminated independently of the main collaboration paper or not. A common model that our research group accomplishes is two manuscripts: one primarily biological where we are second and second-last authors; and the other primarily methodological, where we are first and last authors.

Discuss author order expectations early, particularly for first author, senior and corresponding roles. Also consider co-contributions, co-correspondence, and flexibility for future updates (i.e., new people joining the team). Do not neglect the ordering within co-authors, if applicable. Be aware of possible sources of biases during this negotiation, including gender, PI's favouritism, seniority or tradition (i.e., wet labs taking precedence over dry lab). The same considerations apply to senior and corresponding authors. 

Actionable items: Embrace author contribution taxonomies like CRediT \cite{allen2014publishing} (\url{https://credit.niso.org/}), re-evaluate the policy in case of major change of contributions or because of new collaborators joining while the project is ongoing.

\section*{Rule 4: Jointly design the experiments} 
\label{rule4_experiments}


\begin{flushright}
\rightskip=1cm\textit{``No one believes an hypothesis except its originator but everyone believes an experiment except the~experimenter''} \\
\vspace{.2em}
\rightskip=0cm -- William Ian Beardmore Beveridge
\end{flushright}

Under experimental design, we are referring to the process of defining the general workflow to reach a certain outcome such as testing a hypothesis, answering a biological question or demonstrating a new experimental assay. This encompasses primarily the wet lab experiments but can also include data analysis. Ideally, the project members would define the core aspects of the experimental design prior to collecting data. As a first step, the goal of the research project should be clearly formulated and agreed upon. Both teams should then be involved in the definition of all steps of the experimental design. This may happen at different levels depending on their area of expertise, but all members should have a basic knowledge on all steps of the workflow, including wet lab and dry lab. After defining the experimental design, you should agree with your collaborators on a rough analysis plan. After defining the main steps of the experimental design and analysis plan, a rough timeline should also be generated (see also \nameref{rule5_time}). Finally, the minimal desired outcome (such as initial data analysis, final figures, etc) should also be defined. 

The agreed-upon timeline may be reevaluated after obtaining the initial results. In this case, any changes in the experimental design and/or analysis should be discussed and agreed upon by members of both the wet lab and dry lab. 

Often, dry labs are approached by wet labs to initiate a collaboration (see \nameref{rule1_choose}) after the research question has already been defined and initial wet lab experiments have been performed (and sometimes after primary bioinformatics analysis has been conducted). Starting a collaboration at this stage is often risky and harder than being engaged from the start of the project, since strong expectations for the project may already be set by the wet lab researchers. You should then evaluate whether the proposition from the wet lab users makes sense. If needed, they should suggest additional controls / experiments that would be helpful for downstream data analysis or to reach the final aim of the project.

Actionable items: include test and pilot experiments in your experimental design.

\section*{Rule 5: Agree on project and time management} 
\label{rule5_time}

\begin{flushright}
\rightskip=1cm\textit{``If it's your job to eat a frog, it's best to do it first thing in the morning. And If it's your job to eat two frogs, it's best to eat the biggest one first''} \\
\vspace{.2em}
\rightskip=0cm -- Mark Twain
\end{flushright}

Define together individual tasks and responsibilities, the days or working hours allocated to that project (for communication and urgent tasks to perform), as well as time horizons (in calendar months) for these and for the collaboration more generally. This is also important to get a grasp of the complexity of tasks beyond your expertise that one might underestimate. Plan with buffer time for unforeseen complications. Explicitly check your expectations are aligned (Table~\ref{form}).

To ensure mutual awareness and help monitor progress, it can be useful to keep an up-to-date project plan where everyone can see what was done and what is currently being worked on by whom. If there are digressions or changes of plan, discuss them with the collaborators ahead of engaging in many hours of work. Since most people involved will also have a number of other engagements (and holidays!) with loads that vary over time, it is important to inform each other, ahead of time if possible, of the varying availability for collaboration. Regular meetings, either planned or in response to new data or emerging problems, are also critical to a smooth collaboration (see \nameref{rule6_communication}). 

Actionable items: use online project planners (i.e., Trello, GitHub projects, etc); use multiuser online text editors (i.e., Google Docs, Overleaf) to draft manuscripts and other documents.

\section*{Rule 6: Communicate early openly and often enough} 
\label{rule6_communication}

\begin{flushright}
\rightskip=1cm\textit{``The single biggest problem in communication is the illusion that it has taken place''} \\
\vspace{.2em}
\rightskip=0cm -- George Bernard Shaw
\end{flushright}

Establishing a productive and respectful communication strategy is essential for fruitful and low-friction collaborations. To ensure clarity, a mutual expectation agreement is also needed regarding project length, cost sharing, preferred communication channels and frequency, as well as expected response times. Regular meetings, prompt feedback (for a mutually-agreed definition of prompt), and attending relevant meetings all help maintain alignment on communication. Communication should not only take place when you need something concrete from your collaborators, but sharing and feedback require continuous cycling. However, all collaborators should be empathetic and flexible in meeting plans or project progress due to other responsibilities (teaching duties, other projects, personal reasons) and with mutual respect to others’ work hours (see \nameref{rule5_time}). Especially after long or complicated discussions, or when multiple people were involved, it can be helpful to prepare a summary of important discussion points and decisions to be circulated in written form (e.g., Google Docs or via a Slack or Mattermost channel), to represent a common understanding and have a memento of the conversations.

Actionable items: Adhere to effective meetings \cite{leblanc2019planning} and written interactions guidelines \cite{gruber2020email}.

\section*{Rule 7: Follow open and reproducible science guidelines} 
\label{rule7_repro}

\begin{flushright}
\rightskip=1cm\textit{``The way of the world is to make laws, but follow custom''} \\
\vspace{.2em}
\rightskip=0cm -- Michel de Montaigne
\end{flushright}

Project members should follow reproducible research practices (e.g., use FAIR file formats, version control, free and open source software, good coding practices and reproducible reporting \cite{wilkinson2016fair,sandve2013ten,stawarczyk2023establishing} to ensure that data analysis is transparent and everyone can recreate your results and findings. Before a project starts, agree on the desired degree of public data availability (see also \nameref{rule2_data}) and how (and if) to report summary statistics of sensitive data. Typically, journal policies require authors to describe their data and code availability; this leaves room to share anything from raw data to intermediate results, as well as unstructured code to analysis reports to fully reproducible pipelines. Make sure your collaborators agree with your high commitment to transparency and reproducibility. As a project goes on, it becomes increasingly important to keep all data, software environments, online code, and reports tractable. Ensuring code and results (figures and tables) are in accessible states and versioned enables everyone to stay updated on discussions, failed attempts, and outcomes. Similarly, literate programming \cite{knuth1984literate}, e.g., generating reports with both code and results in open formats (HTML, PDF), facilitates scientific reproducibility and open access. Reproducible computational research is crucial for guaranteeing the consistency and transparency of scientific discoveries. Conversely, a lack of reproducibility threatens all findings.

Since the expertise of both sides differ, it is important to agree on an adequate strategy for sharing results and adapt it to expectations and technical abilities. Although sharing the code among analysts is (or should be) a standard, reports shared with experimentalists could be easier to digest when limited to figures, main results and conclusions. The format to share results should also be clearly defined (meetings, static reports, dynamic reports, etc) and always discussed between the two parts to promote a clear understanding of the collaboration's outputs.   

Actionable items: use version control (git) and push frequently to remotes shared with your collaborators (i.e., GitHub, BitBucket, GitLab); include reproducible and versioned software installation steps in your code repositories; create an analysis workflow that can easily be rerun after a change in the data or in the analysis.

\section*{Rule 8: Establish the transparency and trust required for constructive feedback} 
\label{rule8_trust}

\begin{flushright}
\rightskip=1cm\textit{``Far better an approximate answer to the right question, which is often vague, than an exact answer to the wrong question, which can always be made precise''} \\
\vspace{.2em}
\rightskip=0cm -- John Tukey
\end{flushright}

Project teams should jointly engage in thorough sanity checks throughout the research. This involves regularly questioning and testing various aspects of the project, in both wet and dry lab domains (see~\nameref{rule7_repro}). It might include repeating experiments to confirm their consistency, and using different analysis approaches to explore robustness of results to analysis choices. Both sides should be transparent about the specifics and the output of their work, including unsuccessful attempts, failed controls, irreproducible or not-straightforward-to-interpret results. The reporting of unsuccessful results not only helps to build trust and transparency between both sides, but also helps to discuss alternatives and improve the (wet lab or dry lab) process. Results should be shared fully, e.g., raw western blots, raw imaging data, and all quality control checks. The goal of such a continuous policy of critical thinking and sanity checking is to maintain integrity and build trust among collaborators; and to confirm that both the experimental setups and computational analyses are functioning as expected. By routinely challenging the integrity and interpretability of wet lab and analysis results, potential misalignments can be avoided.

Actionable items: share all raw and processed data; listen to and acknowledge dissenting opinions.

\section*{Rule 9: Be respectful and show appreciation} 
\label{rule9_respect}

\begin{flushright}
\rightskip=1cm\textit{``By mutual confidence and mutual aid great deeds are done, and great discoveries made''} \\
\vspace{.2em}
\rightskip=0cm -- Homer
\end{flushright}

Acknowledge that everyone who has agreed to take part in the collaboration has their validity and represents a different aspect of the scientific endeavour. There is no role that is \textit{per se} more important than the others, no matter how niche it is. Mutual respect of each other’s work and (domain-specific) knowledge should be a given. In turn, this does not rule out but rather includes the questioning and critical assessment of each other's results on a scientific level in case of doubt (see~\nameref{rule8_trust}). Limitations of both scientific approaches but also personal availability for the project should be clearly stated and respected by the other parties involved (see \nameref{rule6_communication}). Polite communication between collaborators is required at all times, be it when discussing potentially unsatisfying results or agreeing on a time schedule for future steps. Be kind: do not forget to regularly acknowledge and thank others for their work, and explicitly remind you and others the project could not be carried out alone.

Actionable items: be respectful; be aware of common cognitive biases, including the impostor syndrome \cite{clance1978imposter} and Dunning–Kruger effect \cite{kruger1999unskilled}.

\section*{Rule 10: Learn as a team} 
\label{rule10_learn}

\begin{flushright}
\rightskip=1cm\textit{``Question everything. Learn something. Answer nothing''} \\
\vspace{.2em}
\rightskip=0cm -- Euripides 
\end{flushright}

The more you know about what the other side is doing and why, the easier the collaboration. It is of great help if everyone involved has both a basic understanding of the different experimental and computational components involved as well as the willingness to gain a deeper understanding of each others methodology as the collaboration proceeds. What can we really interpret from a PCA plot or how does the library preparation protocol influence sequencing results? This common knowledge is not a requirement at the start of a collaboration but can be established during the project. Treat knowledge gaps as an opportunity to learn something new. This should be practised at all levels, irrespective of level of seniority. It will make the common work and communication a lot easier. For example, it will save the analyst a lot of time if the metadata is machine-readable, in turn a wet-lab collaborator might gain more independence if the analysis scripts and outputs are easily accessible to them and well documented. Take advantage of the diversity of backgrounds as it provides different perspectives on a problem that can offer new approaches to solve problems.

Actionable items: add explanatory comments to your analysis’ code; don’t shy away from asking what you don’t understand; commit to answering basic and advanced questions about the analysis.

\section*{Conclusion} 

In any situation when multiple (scientific) cultures collide, there can be tension and misunderstandings at some stage and there is no simple formula to navigate all the personalities and egos involved. Let us not forget that along the way, all members of a collaboration need to absorb a vast assortment of private as well as work pressures, power differentials, deadlines and commitments. Nonetheless, these differences and cultural quirks are a reality to be observed, studied and understood such that we can embrace the diversity of scientific mindsets. We also note that the goal is not necessarily to pursue fully harmonious collaborations, because stumbling through our misunderstandings helps us crystalize norms for collaborations and define our core values of doing science. This may lead to lifelong collaborations or also to those that will never happen again.

Collaborative data analysis is not our primary role, but in many cases becomes the springboard to new methodological projects. The ten simple tips discussed here are largely about being open, organized, empathetic, professional, practical, systematic, and fair.

\section*{Acknowledgments} 

We thank our past collaborators for their often-pleasurable, sometimes-painful contributions that led to to distilling these rules.

The content of this manuscript was brainstormed during a lab retreat in the scenic town of Aeschi bei Spiez, Switzerland.

\section*{Author contributions}

I.M. conceived the project. M.D.R. and I.M. supervised it. All authors drafted, edited and reviewed the manuscript.

\section*{Funding}

This work has not received any specific funding.

\bibliographystyle{apalike}  
\bibliography{ten_simple_rules_collaborations_2024}  

\begin{thebibliography}{}

\bibitem[Allen et~al., 2014]{allen2014publishing}
Allen, L., Scott, J., Brand, A., Hlava, M., and Altman, M. (2014).
\newblock Publishing: Credit where credit is due.
\newblock {\em Nature}, 508(7496):312--313.

\bibitem[Broman and Woo, 2018]{broman2018data}
Broman, K.~W. and Woo, K.~H. (2018).
\newblock Data organization in spreadsheets.
\newblock {\em The American Statistician}, 72(1):2--10.

\bibitem[Clance and Imes, 1978]{clance1978imposter}
Clance, P.~R. and Imes, S.~A. (1978).
\newblock The imposter phenomenon in high achieving women: Dynamics and
  therapeutic intervention.
\newblock {\em Psychotherapy: Theory, research \& practice}, 15(3):241.

\bibitem[{Data~Protection~Working~Party}, 2011]{party2011advice}
{Data~Protection~Working~Party} (2011).
\newblock Advice paper on special categories of data (“sensitive data”).
\newblock {\em Article 29 of Directive 95/46/EC, Ares 444105}.

\bibitem[Gruber et~al., 2020]{gruber2020email}
Gruber, J., L.H., S., and van Bavel, J. (2020).
\newblock A scientist's guide to email etiquette.

\bibitem[Knuth, 1984]{knuth1984literate}
Knuth, D.~E. (1984).
\newblock Literate programming.
\newblock {\em The computer journal}, 27(2):97--111.

\bibitem[Kruger and Dunning, 1999]{kruger1999unskilled}
Kruger, J. and Dunning, D. (1999).
\newblock Unskilled and unaware of it: how difficulties in recognizing one's
  own incompetence lead to inflated self-assessments.
\newblock {\em Journal of personality and social psychology}, 77(6):1121.

\bibitem[LeBlanc and Nosik, 2019]{leblanc2019planning}
LeBlanc, L.~A. and Nosik, M.~R. (2019).
\newblock Planning and leading effective meetings.
\newblock {\em Behavior Analysis in Practice}, 12(3):696--708.

\bibitem[Sandve et~al., 2013]{sandve2013ten}
Sandve, G.~K., Nekrutenko, A., Taylor, J., and Hovig, E. (2013).
\newblock Ten simple rules for reproducible computational research.
\newblock {\em PLoS Computational Biology}, 9(10):e1003285.

\bibitem[Shabani and Marelli, 2019]{shabani2019reidentifiability}
Shabani, M. and Marelli, L. (2019).
\newblock Re-identifiability of genomic data and the gdpr: Assessing the
  re-identifiability of genomic data in light of the eu general data protection
  regulation.
\newblock {\em EMBO reports}, 20(6):e48316.

\bibitem[Stawarczyk and Roos, 2023]{stawarczyk2023establishing}
Stawarczyk, B. and Roos, M. (2023).
\newblock Establishing effective cross-disciplinary collaboration: Combining
  simple rules for reproducible computational research, a good data management
  plan, and good research practice.
\newblock {\em PLoS Computational Biology}, 19(4):e1011052.

\bibitem[Wilkinson et~al., 2016]{wilkinson2016fair}
Wilkinson, M.~D., Dumontier, M., Aalbersberg, I.~J., Appleton, G., Axton, M.,
  Baak, A., Blomberg, N., Boiten, J.-W., da~Silva~Santos, L.~B., Bourne, P.~E.,
  Bouwman, J., Brookes, A.~J., Clark, T., Crosas, M., Dillo, I., Dumon, O.,
  Edmunds, S., Evelo, C.~T., Finkers, R., Gonzalez-Beltran, A., Gray, A.~J.,
  Groth, P., Goble, C., Grethe, J.~S., Heringa, J., 't~Hoen, P.~A., Hooft, R.,
  Kuhn, T., Kok, R., Kok, J., Lusher, S.~J., Martone, M.~E., Mons, A., Packer,
  A.~L., Persson, B., Rocca-Serra, P., Roos, M., van Schaik, R., Sansone,
  S.-A., Schultes, E., Sengstag, T., Slater, T., Strawn, G., Swertz, M.~A.,
  Thompson, M., van~der Lei, J., van Mulligen, E., Velterop, J., Waagmeester,
  A., Wittenburg, P., Wolstencroft, K., Zhao, J., and Mons, B. (2016).
\newblock The fair guiding principles for scientific data management and
  stewardship.
\newblock {\em Scientific Data}, 3(1):160018.

\end{thebibliography}

\clearpage

\appendix

\setlength\extrarowheight{4pt}

\begin{table}[!ht]
   \caption{Collaboration expectations form. The idea here is that there are no good answers (well, sometimes there are), but that each party fills the form and then compares their answer, to hopefully align expectations. Some questions are phrased rhetorically, but meant to explicitly prepare all members for a collaboration's potential tension points.}
    \label{form}
    \raggedleft
    \begin{adjustbox}{max width=\textwidth}
    \begin{tabular}{|p{3in}|l|l|l|l|l|p{3in}|}
    \hline
        ~ & 1 & 2 & 3 & 4 & 5 & ~ \tabularnewline \hline
        \textbf{Choosing collaborators} & ~ & ~ & ~ & ~ & ~ & ~ \tabularnewline \hline
        The questions on this form must be adhered to as agreed & ~ & ~ & ~ & ~ & ~ & Questions on this form are only for orientation \\ \hline
        Scope and whether to continue collaborating should be discussed regularly & ~ & ~ & ~ & ~ & ~ & Once the collaboration has started, it will be seen through to a deliverable, no matter what happens \\ \hline
        \textbf{Experimental design} & ~ & ~ & ~ & ~ & ~ & ~ \\ \hline
        The wet lab people determine the experimental design/strategy, the dry lab people determine the analysis strategy & ~ & ~ & ~ & ~ & ~ & Both experimental and analysis strategies are to be discussed and jointly agreed on by both parties \\ \hline
        Analysis plan and experimental design are to be discussed once the data has been collected & ~ & ~ & ~ & ~ & ~ & Analysis plan and experimental design should be discussed and fixed prior to data collection \\ \hline
        Analysts are blinded whenever possible & ~ & ~ & ~ & ~ & ~ & All metadata is always visible to analysts \\ \hline
        \textbf{Communication and functioning} & ~ & ~ & ~ & ~ & ~ & ~ \\ \hline
        Meetings should be held on a regular (e.g. monthly/biweekly) basis, and simply be brief if there was no development & ~ & ~ & ~ & ~ & ~ & Meetings should be organized whenever they are needed (e.g. to discuss results or problems) \\ \hline
        Both parties should always adhere to agreed deadlines & ~ & ~ & ~ & ~ & ~ & Agreed deadlines are guidelines rather than absolute targets \\ \hline
        To each his/her expertise & ~ & ~ & ~ & ~ & ~ & All parties are expected to try to learn and understand what the others are doing \\ \hline
        As much as possible, every piece of data/analysis should be made available to everyone & ~ & ~ & ~ & ~ & ~ & Results are to be exchanged/communicated between partners in the form of publication-ready figures and descriptions \\ \hline
        Emails are expected to be answered on the same day & ~ & ~ & ~ & ~ & ~ & Emails are expected to be answered within a week or two \\ \hline
        Both parties agree on tools to track progress (i.e. Trello) and collaborative writing (i.e. Google Docs, Overleaf) & ~ & ~ & ~ & ~ & ~ & Each team uses their preferred software \\ \hline
        Predefined file naming conventions & ~ & ~ & ~ & ~ & ~ & No convention but file names are clear \\ \hline
        Metadata delivered in machine readable format & ~ & ~ & ~ & ~ & ~ & Metadata sent as is collected \\ \hline
        Manuscript written in a collaborative document allowing concurrent editing & ~ & ~ & ~ & ~ & ~ & Different versions of manuscripts are individual documents shared by email \\ \hline
        \textbf{Scientific values and publication strategy} & ~ & ~ & ~ & ~ & ~ & ~ \\ \hline
        What matters is excellent science - and hence high impact factor & ~ & ~ & ~ & ~ & ~ & What matters is sound, robust science \\ \hline
        We should make code/data available if we are forced to & ~ & ~ & ~ & ~ & ~ & By default everything should be openly available, latest by publication time (whether pre-print or final paper) \\ \hline
        Everything should be published as pre-print & ~ & ~ & ~ & ~ & ~ & Never mention anything unless it’s actually published by a journal \\ \hline
        The people doing the experiments are first authors, the dry lab people are somewhere after that & ~ & ~ & ~ & ~ & ~ & The people analyzing and making sense of the data are first authors or first co-authors \\ \hline
        Negative or inconclusive results are sadly unglamorous, but happen and are part of science & ~ & ~ & ~ & ~ & ~ & If you have negative or inconclusive results, it means you haven't looked hard enough or given it proper attention \\ \hline
        \textbf{Intellectual contribution and attribution} & ~ & ~ & ~ & ~ & ~ & ~ \\ \hline
        Projects are owned and conceived by the wet lab & ~ & ~ & ~ & ~ & ~ & Projects are owned and conceived by the dry lab \\ \hline
        Results are interpreted by the dry lab & ~ & ~ & ~ & ~ & ~ & Results are interpreted by the wet lab \\ \hline
        Figure making is a graphic design task (using an image editor) & ~ & ~ & ~ & ~ & ~ & Figure making is a data analysis task (using ggplot, GraphPad, etc) \\ \hline
        \textbf{Etiquette} & ~ & ~ & ~ & ~ & ~ & ~ \\ \hline
        Direct communication is not harsh, just direct & ~ & ~ & ~ & ~ & ~ & Well thought, polite phrasing is expected in all communication exchanges \\ \hline
        Excluding collaboration members from some exchanges (meetings, messages) is ok & ~ & ~ & ~ & ~ & ~ & All exchanges should be public and no member can be excluded \\ \hline
        Freedom to give feedback regardless of career stage & ~ & ~ & ~ & ~ & ~ & Mind status when giving feedback \\ \hline
    \end{tabular}

    \end{adjustbox}
\end{table}

\end{document}